\documentclass[prl,twocolumn,showpacs,amsmath,superscriptaddress,psfig,amssymb]{revtex4}
\usepackage{graphicx}% Include figure files
\usepackage{dcolumn}% Align table columns on decimal point
\usepackage{bm}% bold math

\begin{document}

%\preprint{APS/123-QED}

\title{Revised Phase Diagram for FeTe$_{1-x}$Se$_{x}$ system with less excess Fe atoms}

\author{Chihen Dong}
\affiliation {Department of Physics, Zhejiang University, Hangzhou, Zhejiang 310027, China}

\author{Hangdong Wang}
\affiliation {Department of Physics, Zhejiang University, Hangzhou 310027, China}
\affiliation {Department of Physics, Hangzhou Normal University, Hangzhou, Zhejiang 310036, China}

\author{Zujuan Li}
\affiliation {Department of Physics, Zhejiang University, Hangzhou 310027, China}

\author{Jian Chen}
\affiliation {Department of Physics, Zhejiang University, Hangzhou 310027, China}

\author{H. Q. Yuan}
\affiliation {Department of Physics, Zhejiang University, Hangzhou 310027, China}

\author{Minghu Fang}
\email{mhfang@zju.edu.cn}
\affiliation {Department of Physics, Zhejiang University, Hangzhou 310027, China}

\date{\today}% It is always \today, today,
             %  but any date may be explicitly specified

\begin{abstract}
\noindent We observed bulk superconductivity in the FeTe$_{1-x}$Se$_{x}$ crystals with a low Se concentration after the excess Fe atoms existing unavoidably in crystals as-grown are partially removed by means of annealing in the air. A revised magnetism and superconductivity phase diagram is obtained via resistivity and magnetic susceptibility measurements for FeTe$_{1-x}$Se$_{x}$ system. It is found that bulk superconductivity coexists with antiferromagnetic order in the crystals with 0.05$<$x$<$0.20. The phase diagram is very similar to the case of the K-doping or Co-doping BaFe$_{2}$As$_2$, as well as of SmFeAsO$_{1-x}$F$_{x}$ iron-pnictide system, perhaps indicating that all iron-based systems have a generic phase diagram, although antiferromagnetic wave vector in the parent compounds of these systems is remarkable different.
\end{abstract}

% PACS, the Physics and Astronomy
                             % Classification Scheme.
\maketitle

The discovery of superconductivity in the iron pnictides has generated much interest in understanding the interplay of magnetism and superconductivity in iron-based layered compounds\cite{Kamihara 2008,Chen 2008}. There are, so far, five different structural classes of Fe-based compounds reported, \textit{i.e.}, 1111-type ReOFeAs (Re=rare earth)\cite{GFChen 2008,Ren 2008,Ren 2008-1,Cheng 2008,Wang 2008}, 122-type AFe$_{2}$As$_{2}$ (A=Ba, Sr, or Ca)\cite{Rotter 2008,Sefat 2008}, 111-type BFeAs (B=alkali metal)\cite{Wang 2008-1,Tapp 2008,Pittcher 2008}, 11-type Fe(Te, Se, S)\cite{Hsu 2008,Fang 2008,Mizuguchi 2009,Dong 2010} and 32522-type Sr$_{3}$Sc$_{2}$O$_{5}$Fe$_{2}$As$_{2}$ \cite{Zhu 2009,Han 2010,Kotegawa 2009}. Their magnetic and superconducting phase diagrams are qualitatively similar to those of several other classes of unconventional superconductors, including the cuprates \cite{Damascelli 2003,Armitage 2010}, organics \cite{Lefebvre 2000,Powell 2005} and heavy-fermion superconductors \cite{Park 2006,Yuan 2003}. In all cases antiferromagnetism (AFM) must be suppressed, either by doping or pressure, before optimal bulk superconductivity (SC) appears. But, up to now, it has also been found that there are somewhat differences in the details of the phase diagram among the Fe-based systems. For fluorine-doped '1111' systems such as LaFeAsO$_{1-x}$F$_{x}$ \cite{Luetkens 2009}, AFM and SC phases are completely separated as a function of doping and do not overlap. In BaFe$_{2}$As$_{2}$, the systematic substitution of either the alkaline-earth (Ba) \cite{HChen 2009,Rotter 2008-1}, transition-metal (Fe) \cite{Nandi 2010,Sharma 2010} or pnictogen (As) \cite{Jiang 2009} atom with a different element almost universally produces the phase diagram, which is composed of coupled AFM and structural transitions that are suppressed with substitution and an SC phase that is more or less centered near the critical concentration where AFM order is destroyed. The phase diagram obtained from the Fe$_{1+y}$Te$_{1-x}$Se$_{x}$ system \cite{Khasanov 2009,Liu 2010,Katayama 2010}, up to now, is somewhat different from that in 1111 and 122-type systems, containing a spin glass phase between the AFM and SC phases. Long-range AFM ordering ceases at x=0.1, and bulk SC does not set in until x exceeds 0.3 to 0.4.

For Fe$_{1+y}$Te$_{1-x}$Se$_{x}$ system, there is another important difference than the Fe-pnictide (1111- and 122-type) system. The parent compounds of Fe-pnictide system are in a collinear AMF order, which is characterized by the in-plane Fermi surface nesting wave-vector \textit{Q}$_n$=($\pi$,$\pi$) \cite{Luetkens 2009,Clarina 2008,Huang 2008}, (here we refer to wave vectors in units of the inverse tetragonal lattice parameters). But, Fe$_{1+y}$Te, the parent compound of the Fe$_{1+y}$Te$_{1-x}$Se$_{x}$ superconductors, is in a bi-collinear AFM order, which is characterized by an in-plane wave vector \textit{Q}$_m$=($\pi$,0) \cite{Bao 2009, Li 2009-1}, although both doped iron-pnictide \cite{Christianson 2008, Lumsden 2008, Chi 2009, Li 2009-2} and iron-chacolgenide \cite{Qiu 2010} superconductors exhibit a magnetic response around the same wave vector ($\pi$,0) in the spin excitation spectra below superconducting transition temperature, \textit{T}$_C$, consistent with \textit{S}$_{\pm}$ state that emerges as a likely candidate pairing state for Fe-based superconductors. Thus, it is quite natural to suggest \cite{Liu 2010, Katayama 2010} that there is an intermediate region without bulk superconductivity with magnetic correlations near ($\pi$,0), which are not favorable to superconductivity, in the phase diagram, between long-range ($\pi$,0) magnetic order phase and SC phase with ($\pi$,$\pi$) magnetic resonance. However, in the Fe$_{1+y}$Te$_{1-x}$Se$_{x}$ system, the presence of excess Fe atoms makes a complicated magnetism \cite{Bao 2009} and may provide magnetic pair breaking effect\cite{Zhang 2009}. In fact, the phase diagram of Fe$_{1+y}$Te$_{1-x}$Se$_{x}$ system discussed above may include the factor of the excess Fe atoms. In this letter, we removed partially the excess Fe atoms existing unavoidably in the crystals as-grown by annealing in the air, and found that there is no intermediate region in the phase diagram for FeTe$_{1-x}$Se$_{x}$ system with less excess Fe atoms. Instead, SC phase with ($\pi$,$\pi$) magnetic resonance can coexist with the long-range ($\pi$,0) magnetic order phase in a rather wide Se concentration range. The surprisingly similarity of the phase diagram of FeTe$_{1-x}$Se$_{x}$system with less excess Fe atoms to that in the doping BaFe$_{2}$As$_{2}$ system, as well as that in SmFeAsO$_{1-x}$F$_{x}$ reported recently \cite{Drew 2009}, indicate that the coexistence of AFM and SC phases is probably the more intrinsic property of the generic Fe-based superconductors phase diagram.

Single crystals of Fe$_{1+y}$Te$_{1-x}$Se$_{x}$ were grown using the Bridgeman technique. A mixture of Fe, Te, and Se powders with the appropriate ratios (we choose y=0 for decreasing the content of excess Fe atoms) was heated to 920$^o$C in an evacuated tube then slowly cooled, forming single crystals as-grown. In order to remove the excess Fe atoms existing unavoidably in the crystals as-grown, we try to re-heat the crystals in the air (A) at different temperatures ($\leq$300$^o$C) for 2 hours, in vacuum (V) at 400$^o$C for 7 days, respectively, and in both. The composition of crystals as-grown, and annealed in the air and vacuum were determined using energy dispersive X-ray spectrometer (EDXS). The resistance, Hall effects and DC magnetic susceptibility measurements were carried out using commercial Physical Property Measurement System (PPMS), and SQUID, respectively.

Figure 1(a) and (b) show the \textit{ab}-plane normalized resistance (due to the difficulty to determine the exact path of current in the crystals, we use a normalized resistance for comparability),\emph{R}(T), and bulk magnetic susceptibility data, $\chi$(T), for single crystals annealed at 270$^o$C for 2 hours with different Se concentrations, FeTe$_{1-x}$Se$_{x}$ (0.0$\leq$x$\leq$0.4). For x=0, and 0.02 crystals, a sharp decrease in resistance are observed at \textit{T}$_N$=70, 62.6 K, respectively, corresponding to the AFM phase transition, companied with a structural transition, as reported in our original papers \cite{Fang 2008, Bao 2009}, but no drop corresponding to SC transition below \textit{T}$_N$ is observed above 2K. For \textit{x}=0.05, 0.07, and 0.1 crystals, except for a sharp decrease in resistance at \textit{T}$_N$, which decreases with increasing Se concentration, another sharp drop corresponding to SC transition is observed at T$_C$(mid.)=6.7, 8.2, and 9.6K, respectively, zero resistance is obtained at \textit{T}$_C$$^{zero}$=4.0 and 7.6K for \textit{x}=0.07, 0.1 crystals, but no zero resistance obtained for x=0.05 above 2K. To check whether superconductivity in these low Se concentration crystals annealed in the air is a bulk property or not, we measure their bulk susceptibility, as shown in the left inset in Fig. 1(b), presenting an evidence for bulk SC. For the 0.12$\leq$x$\leq$0.4 crystals, both the resistance and susceptibility data show that bulk superconductivity exists in these crystals, although the SC transition in the $\chi$(T) data is not so sharp, and SC volume fraction is smaller than that in the best crystal with x=0.4 annealed in both the air and vacuum. It is remarkable that the temperature dependence of resistance exhibits a semiconductor-like behavior above the SC transition, i.e. its resistance increases with decreasing temperature. Although there is no drop in resistance corresponding to an AFM transition at \textit{T}$_N$ for the crystals with 0.12$\leq$x$\leq$0.18, a drop in the bulk susceptibility measured at 1 Tesla with field cooling process confirms this transition, as shown in the left inset of Fig. 1(a).

The result of the coexistence of bulk SC and long-range AFM order in the crystals annealed in the air with low Se concentrations, is sharply inconsistent with those reported previous \cite{Khasanov 2009,Liu 2010,Katayama 2010}. We argue that the difference in the number of the excess Fe atoms existing unavoidably in the crystals with low Se concentration results in these disagreements. We found also that the crystals either as-grown or annealed in vacuum with \textit{x}$\leq$0.3 indeed do not show bulk SC. It is found that the excess Fe atoms is easily removed partially by annealing in the air, although it is difficulty to confirm the excess Fe atoms in the crystals as-grown completely being removed. Our EDXS data indicate the Fe content of the crystals annealed in the air is less than that in the crystals as-grown or annealed only in vacuum, as discussed in following.

\begin{figure}
  % Requires \usepackage{graphicx}
  \includegraphics[width=8cm]{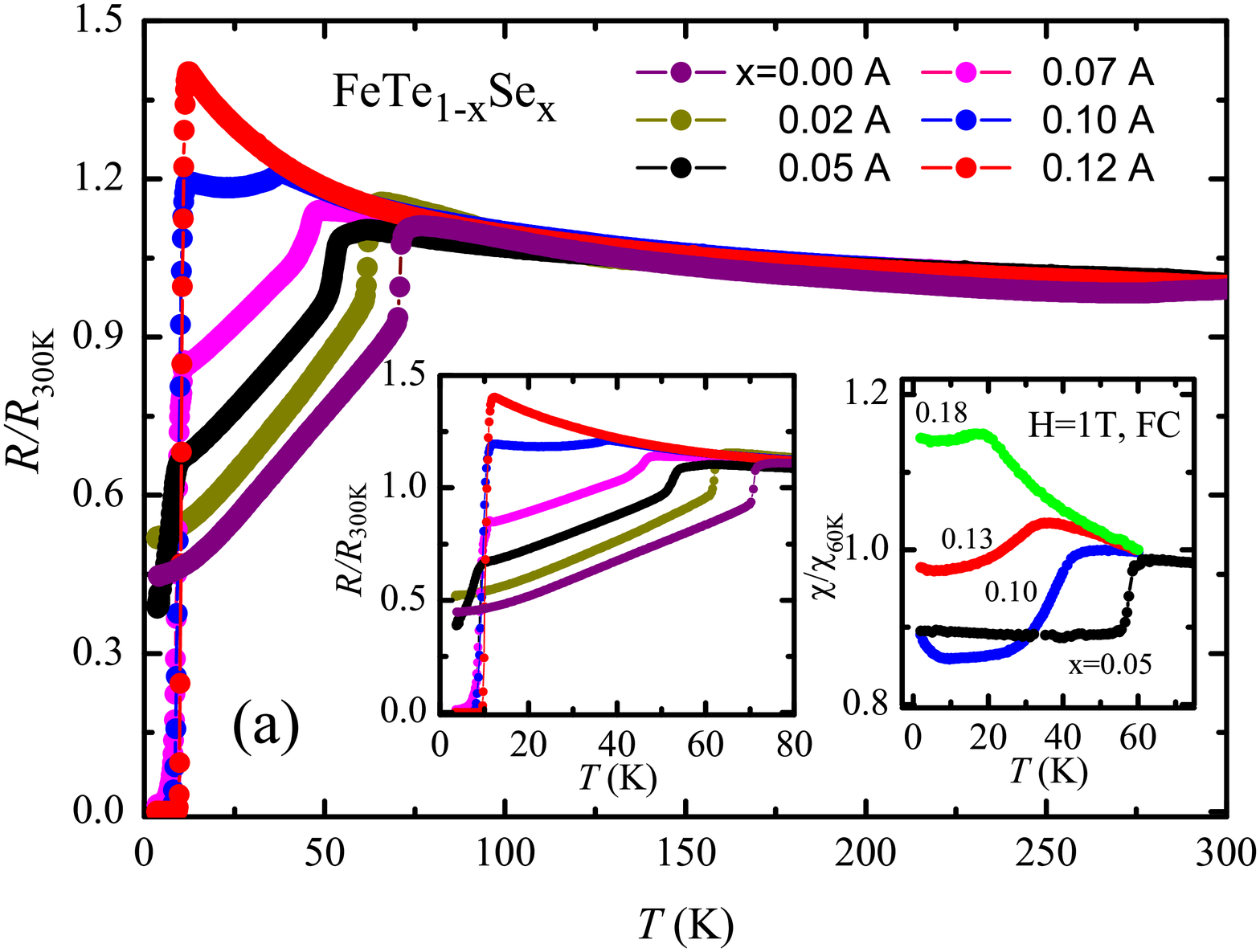}\\
  \includegraphics[width=8cm]{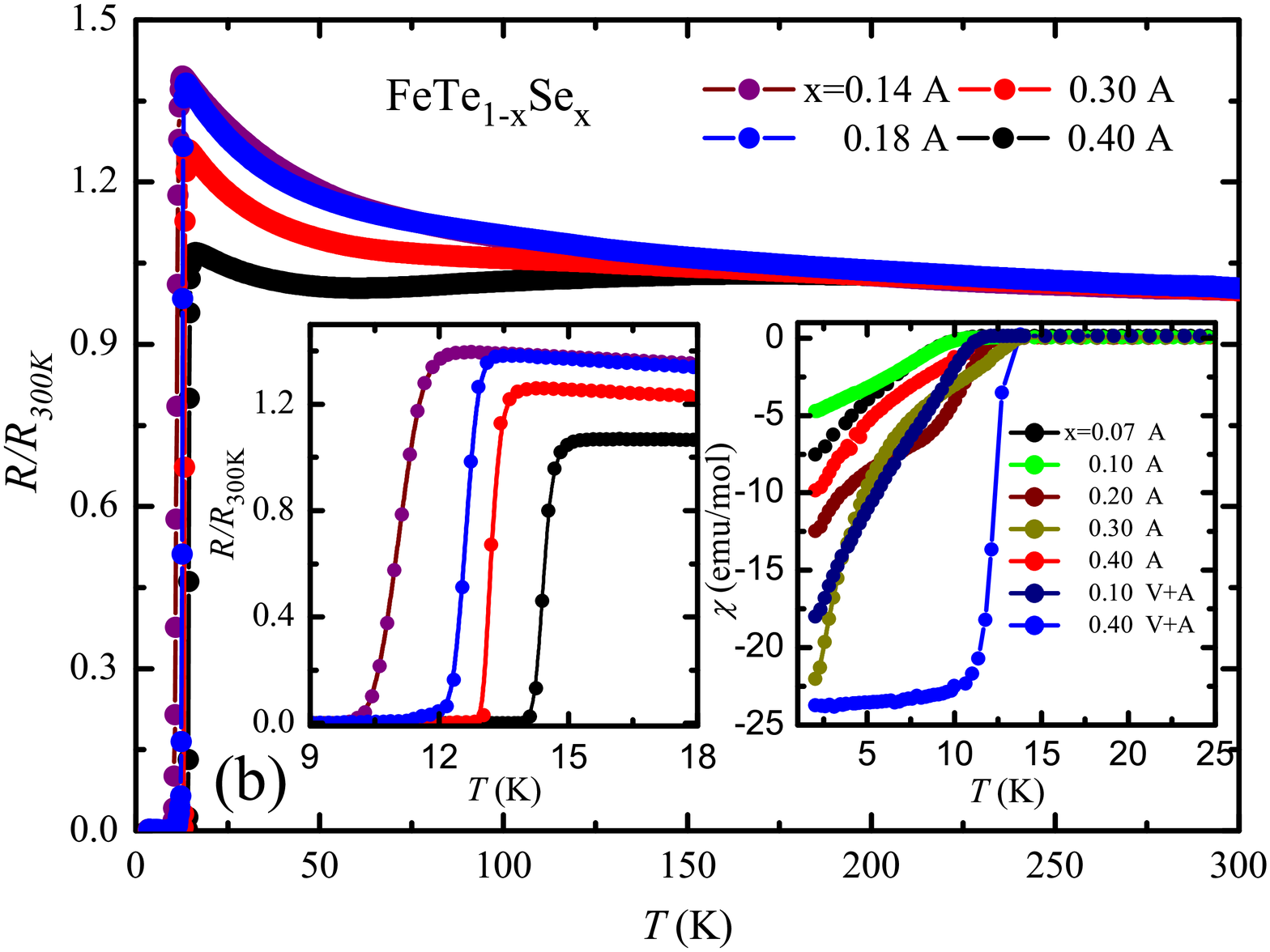}\\
  \caption{(Color online)Temperature dependence of the normalized resistance, \textit{R}$_{ab}$(T)/\textit{R}$_{ab}$(300K), for FeTe$_{1-x}$Se$_{x}$ crystals annealed at 270$^o$C in the air(A). (a)\textit{x}=0.00, 0.02, 0.05, 0.07, 0.10 and 0.12; and the right inset plots the AMF and superconducting transition. (b)\textit{x}=0.14, 0.18, 0.30, and 0.40; and and the right inset plot the superconducting transition. The left inset in (a): temperature dependence of the normal susceptibility measured at 1Tesla with field cooling process, $\chi$(T)/$\chi$(60K) for the \textit{x}=0.05, 0.10, 0.13 and 0.18. The left inset in (b): temperature dependence of the susceptibility, $\chi$(T), measured at 30Oe with field cooling process for \textit{x}=0.07, 0.10, 0.20, 0.30, and 0.40}\label{}
\end{figure}

Figure 2 shows the \textit{x}-T phase diagram based on the resistance and susceptibility data for the FeTe$_{1-x}$Se$_{x}$ crystals annealed in the air at 270$^o$C for 2 hours. The value of \textit{x} is a nominal value and may not be exactly correct. The EDXS data show that the Se concentration value \textit{x} is almost the same as the nominal value, and the Fe concentration value in the crystals annealed in the air is closer to unity than that in crystals as grown or annealed in vacuum. The \textit{T}$_N$ value is determined by the temperature at which the both \textit{dR(T)/dT} and \textit{d$\chi$(T)/dT} have a maximum value. The \textit{T$_C$} value is determined by the middle temperature of the superconducting transition in \textit{R}(T) curves. The phase diagram clearly shows that there is a region (0.05$\leq$x$\leq$0.18) where the AFM phase coexists with bulk SC phase, although the superconducting volume fraction is less than 1 and the superconducting transition in $\chi$(T) is not sharp, indicating that inhomogeneous in superconductivity exists in the crystals. Only near the end member, FeTe, does not show bulk SC, which is the same as that reported in our original paper \cite{Fang 2008}. Our phase diagram of FeTe$_{1-x}$Se$_{x}$ with less excess Fe atoms is very similar to that for K-doping or Co-doping 122 Fe-pnictide system, but much different than that for the same Fe$_{1+y}$Te$_{1-x}$Se$_{x}$ system reported in Refs.\cite{Khasanov 2009, Liu 2010, Katayama 2010}. It is most surprised that the bulk SC exhibiting a magnetic response around the wave vector ($\pi$,$\pi$) coexists with the AFM order characterized by an in-plane wave vector \textit{Q}$_m$=($\pi$, 0). Our finding of the similarity of the phase diagram of FeTe$_{1-x}$Se$_{x}$ system to that in the K- or Co-doing 122-type system, as well as that in SmFeAsO$_{1-x}$F$_{x}$ (1111-type) reported recently \cite{Drew 2009}, indicate that the coexistence of AFM and SC phases may be an intrinsic property of the Fe-based superconductors.

\begin{figure}
  % Requires \usepackage{graphicx}
  \includegraphics[width=8cm]{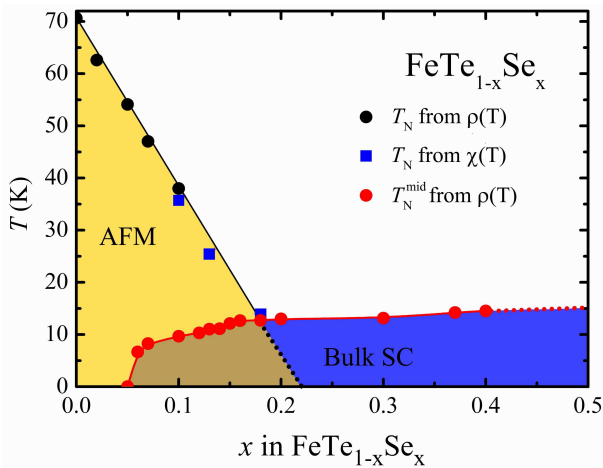}\\
  \caption{(Color online)Phase diagram showing the AFM transition temperature, \textit{T}$_{N}$ and the superconducting temperature \textit{T}$_C$ as function of \textit{x} for FeTe$_{1-x}$Se$_x$ system.}\label{}
\end{figure}

We choose crystals as-grown with a lower (\textit{x}=0.1) and a higher (\textit{x}=0.4) Se concentration to investigate the effects of annealing on SC and magnetism. Figure 3(a) presents the temperature dependence of \textit{ab}-plane normalized resistance, \textit{R$_{ab}$}(T), and DC susceptibility, $\chi$(T), measured at 30Oe ($\parallel$c axis) with zero field cooling process (ZFC) for FeTe$_{0.9}$Se$_{0.1}$ crystals as-grown, annealed in the air (A) at various temperatures for 2 hours, in vacuum (V) at 400$^o$C for 7 days, and first in the air, then in vacuum (A+V), respectively. For the Fe$_{1+y}$Te$_{0.9}$Se$_{0.1}$ crystals as-grown and annealed only in vacuum, as shown in the insets in Fig.3(a), their resistivity and susceptibility data do not show any superconducting transition above 2K. But all the FeTe$_{0.9}$Se$_{0.1}$ crystals annealed in the air at different temperatures, and annealed first in the air then in vacuum exhibit bulk SC below about 9K. And we found that the superconducting volume fraction increases with increasing annealing temperature [see the right inset of Fig.3 (a)], even the superconducting volume fraction in the crystal annealed in the air at 300$^o$C is larger than that annealed in the air at 270$^o$C, then in vacuum, although the latter has a higher \textit{T}$_C$ and sharper superconducting transition in \textit{R}$_{ab}$(T). Another, there is a kink at about \textit{T}$_N$=38K, corresponding to an AFM transition, in the \textit{R}$_{ab}$(T) curves for the all crystals. We found that annealing in the air and vacuum has little effect on \textit{R}$_{ab}$(T) behavior above \textit{T}$_N$, and results in a slight decrease in \textit{T}$_N$ (38K for the crystals annealed in the air at 270$^o$C, while 42K for the crystals as-grown). The \textit{R}$_{ab}$(T) above \textit{T}$_N$ exhibits a semiconducting behavior for all the crystals. The result of the coexistence of bulk SC and an AFM order in FeTe$_{0.9}$Se$_{0.1}$ crystals annealed in the air is sharply inconsistent with that reported in Refs \cite{Liu 2010,Katayama 2010}. To check the effect of annealing in air on the composition of crystals, we measured the composition of the crystals as-grown, annealed in the air at 270$^o$C using EDXS. We found that Fe concentration decreases from 1.02 in the crystals as-grown to almost 1.00 in the crystals annealed in the air, but there is little change in the relative ratio of Te and Se before and after annealed. Although we can not determine the exact composition in crystals using EDXS, the existence of Fe oxides (changing to red brown from black color) on the crystals after annealed in the air is another evidence for the excess Fe atoms being removed by annealing, in contrast, the annealing in vacuum can not remove the excess Fe atoms in the crystals as-grown.

For the higher Se concentration crystals FeTe$_{0.6}$Se$_{0.4}$ as-grown, which is believed \cite{Sales 2009} to have smaller mount excess Fe atoms than crystals with the lower Se concentration and exhibit bulk SC, we found that annealing in the air can also improves its superconductivity. Figure 2(b) presents the temperature dependence of \textit{ab}-plane normalized resistance, \textit{R}$_{ab}$(T), and DC susceptibility, \textit{до}(T) for FeTe$_{0.6}$Se$_{0.4}$ crystals as-grown, and annealed. For the crystals as-grown, although its \textit{R}$_{ab}$(T) data shows a sharp superconducting transition with \textit{T}$_C$$^{onset}$=13K, \textit{T}$_C$$^{zero}$=10K, the superconducting volume fraction is very small, as shown in the right inset in Fig.2(b), and its \textit{R}$_{ab}$(T) at the normal state exhibits a semiconducting behavior. For the crystal annealed only in vacuum, both the \textit{R}$_{ab}$(T) and $\chi$(T) data show that the crystal exhibits bulk SC with \textit{T}$_C$$^{onset}$=13.7K, \textit{T}$_C$$^{zero}$=11.3K, the \textit{T}$_C$ value and superconducting volume fraction are higher than that in the crystals as-grown, and \textit{R}$_{ab}$(T) at the normal state shows a metal behavior. The annealing in vacuum can improve not only SC but also transport properties at the normal state, indicating that the disorders in the crystals, which may result in the localization of carriers, is another factor to suppress SC except for the magnetic pair-breaking effect resulting from excess Fe atoms in the Fe$_{1+y}$Te$_{1-x}$Se$_x$ system.

For all the FeTe$_{0.6}$Se$_{0.4}$ crystals annealed in air, although their \textit{R}$_{ab}$(T) at the normal state still shows a semiconducting behavior as that in the crystal as-grown, their superconductivity is remarkably improved, and the \textit{T}$_C$ and superconducting volume fraction increase with increasing annealing temperature. These results indicate that the existence of excess Fe atoms is the main factor to suppress SC, the localization of carriers due to the disorders in crystals results in the semiconducting behavior at the normal state and has a little effect on SC. The \textit{R}$_{ab}$(T) and \textit{$\chi$}(T) data for the crystals annealed first in the air, then in vacuum present an evidence for this scenario. The crystal annealed in both the air and vacuum has the highest \textit{T}$_C$$^{mid}$=15.2K with the sharpest transition, the largest superconducting volume fraction, and its \textit{R}$_{ab}$(T) at the normal state exhibits also a metallic behavior.

\begin{figure}
  % Requires \usepackage{graphicx}
  \includegraphics[width=8cm]{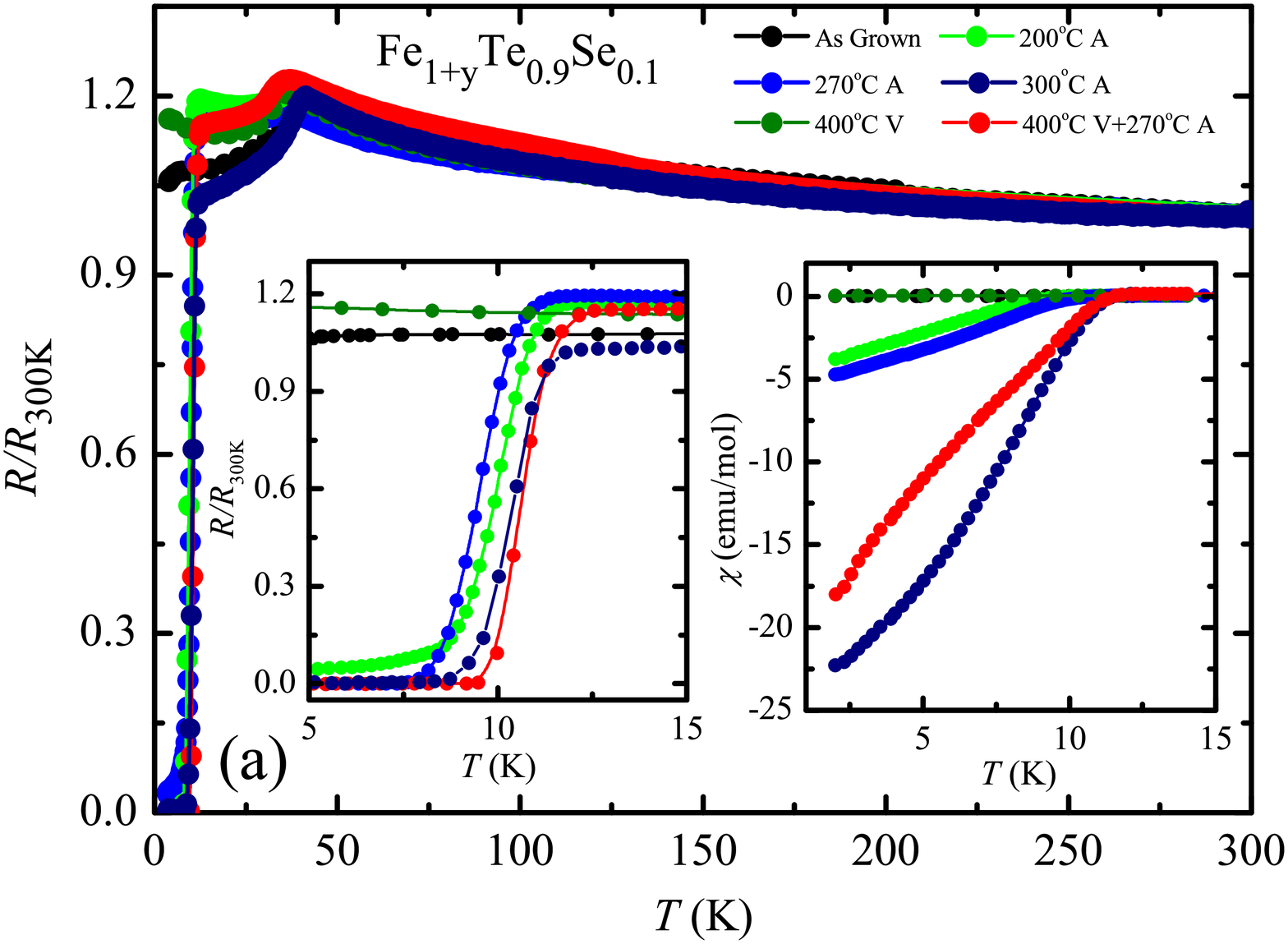}\\
  \includegraphics[width=8cm]{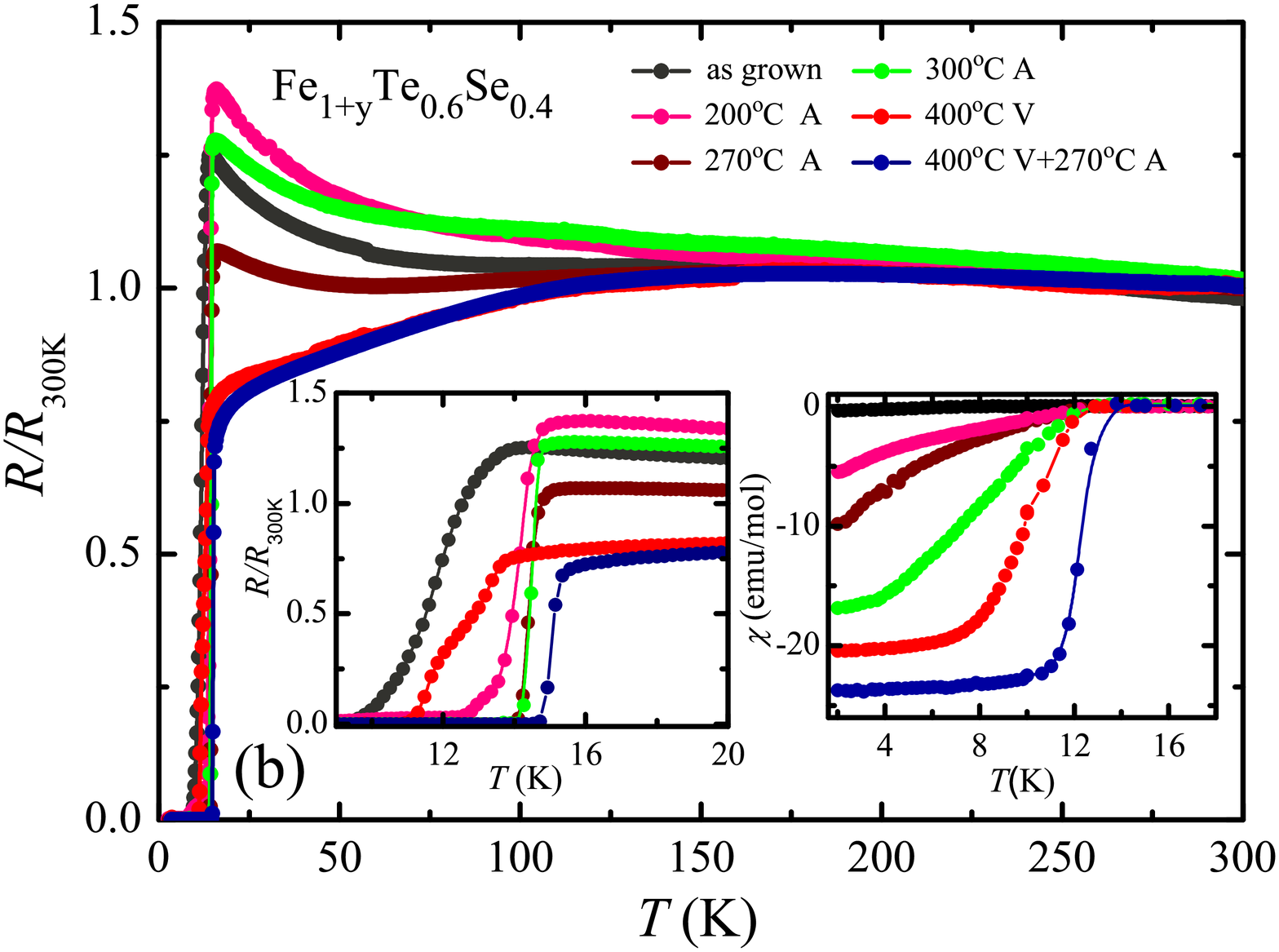}\\
  \caption{(Color online) (a) Temperature dependence of the normalized resistance, \textit{R}$_{ab}$(T)/\textit{R}$_{ab}$(300K) (main plot), the right inset plots the superconducting transition, the left inset plots the susceptibility as a function of temperature, $\chi$(T) measured at 30Oe with zero field cooling for FeTe$_{0.9}$Se$_{0.1}$ crystals as-grown and annealed at different temperature in the air (A) and in vacuum (V). (b) \textit{R}$_{ab}$(T)/\textit{R}$_{ab}$(300K), (main plot), the right inset plots the superconducting transition, the left inset plots $\chi$(T) for FeTe$_{0.6}$Se$_{0.4}$ as-grown and annealed at different temperature in the air and in vacuum.}\label{}
\end{figure}

In order to check the effect of annealing on the electronic structure, we measured Hall coefficients for the FeTe$_{0.6}$Se$_{0.4}$ crystals as-grown, annealed in air and vacuum, respectively, and annealed in both. Figure 4 presents Hall coefficient, \textit{R}$_H$(T), as a function of temperature for FeTe$_{0.6}$Se$_{0.4}$ crystals as-grown and annealed, as well as the crystal as-grown of parent compound, Fe$_{1.02}$Te for comparison. The calculation using the local density approximation (LDA) \cite{Zhang 2009} and experiments by angle-resolved photoemission spectroscopy (ARPES) \cite{Xia 2009} have confirmed that the band structure and Fermi surfaces of FeTe$_{1-x}$Se$_x$ are similar to those of the 122- and 1111-type compounds, and at Fermi energy, there are circular hole pockets at the $\Gamma$ =(0,0) point of the tetragonal Brillouin zone and elliptical electron pockets at the $M$=($\pi$,$\pi$) point. As expected for an undoped semimetal, the numbers of electron and hole carriers were found to be about the same. The Hall coefficient data of Fe$_{1.02}$Te crystal as-grown shown in the inset in Fig. 4 indicate that the AFM transition is accompanied by a remarkable change of the Fermi surface, the Fermi surface is dominated by holes above \textit{T}$_N$ and by electrons below \textit{T}$_N$. For the Fe$_{1+y}$Te$_0.6$Se$_0.4$ crystal as-grown, above 65K, its \textit{R}$_H$ value is almost the same as that in undoped Fe$_{1.02}$Te, and independent of temperature, indicating that the partial substitution of Se for Te does not change the carrier density, which is consistent with the fact of Se and Te having the same valance, and only changes the Fermi surface. The negative peak at near \textit{T}$_C$ in \textit{R}$_H$(T) curve is associated with the vortex motion, which is not a theme discussed here. We focus only on the \textit{R}$_H$(T) behavior at the normal state. The positive value of \textit{R}$_H$ at the normal state indicates that the carrier is dominated by holes. For the crystals annealed either only in vacuum or in both the air and vacuum, their \textit{R}$_H$ value above 60K is almost the same as that of crystals as-grown, but the \textit{R}$_H$ value below 60K has a remarkable decrease, indicating that the annealing in vacuum results in a remarkable increase of the carrier concentration, which is consistent with the scenario that the disorder existing in the crystals as-grown results in a localization of carriers and annealing in the vacuum can remove these disorders. In contrast, the annealing in the air has more notable and complicated effects on the \textit{R}$_H$ (T) behavior. For the crystals annealed at 270$^o$C in the air, their \textit{R}$_H$ value increases with decreasing temperature, especial at low temperatures ($<$30K) the \textit{R}$_H$ values are larger than that of the crystal as-grown, indicating that the charge carrier concentration decreases after the excess Fe atoms are removed. This is consistent with the fact that there is a little change in \textit{R}(T) behavior at normal state, which shows a semiconducting behavior. The enhancement of superconductivity after the excess Fe atoms being removed indicates that suppression of SC due to the existence of the excess Fe atom may result from its magnetic pair-breaking effect, instead of localization of carriers, which is consistent with the calculations using LDA that the excess Fe atoms are magnetic \cite{Zhang 2009}.

\begin{figure}
  % Requires \usepackage{graphicx}
  \includegraphics[width=8cm]{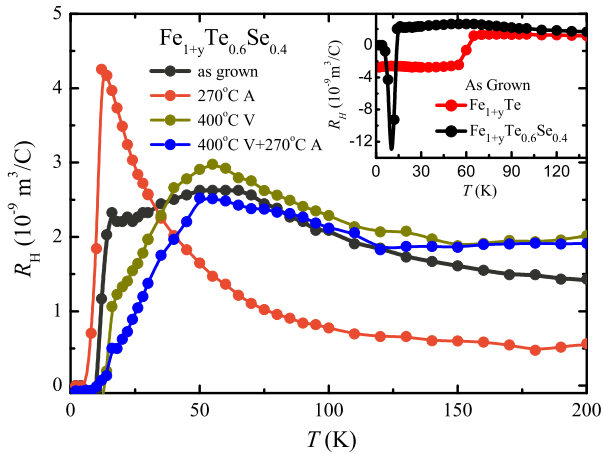}\\
  \caption{(Color online) Temperature dependence of the Hall coefficient for FeTe$_{0.6}$Se$_{0.4}$ crystals as-grown, and annealed in the air, in vacuum and in both (main plot). And for the Fe$_{1.02}$Te as-grown for comparison (upper inset).}\label{}
\end{figure}

In summary, in the FeTe$_{1-x}$Se$_x$ system, we found that the excess Fe atoms existing unavoidably in the crystals as-gown is easily removed by means of annealing in the air. Bulk superconductivity was observed in the low Se concentration crystals after removing partially the excess Fe atoms. A revised magnetism and superconductivity phase diagram for FeTe$_{1-x}$Se$_x$ system with less excess Fe atoms is obtained via resistivity and magnetic susceptibility measurements. It was found that bulk SC coexists with AFM order in the crystals with 0.05$<$x$<$0.20, which is very similar to the case of the K-doping or Co-doping 122 iron-pnictide system, indicating that all iron-based systems have the similar phase diagram, although AFM wave vector in the parent compounds of both systems is remarkable different.

This work is supported by the National Science Foundation of China (Grant No.10974175, 10934005), the National Basic Research Program of China (Grant No.2009CB929104, 2011CB605903), the PCSIRT of the Ministry of Education of China (Contract No.IRT0754).

\end{document}